\documentclass[sigconf]{acmart}
\usepackage{amsmath}
\usepackage{amssymb}             %数学公式
\usepackage{amsfonts}            %数学字体
\usepackage{mathrsfs}            %数学花体
\usepackage{enumerate}
\usepackage{indentfirst}
\usepackage{graphicx}
\usepackage{multirow}
\usepackage{subfigure}
\usepackage{caption}
\usepackage{bm}
\usepackage{booktabs}
\usepackage[linesnumbered,boxed,ruled,commentsnumbered]{algorithm2e}

\begin{document}
\title{Neural Collaborative Ranking}
\author{Bo Song}
\affiliation{Zhejiang University}
\email{bosong16@zju.edu.cn}
\author{Xin Yang}
\affiliation{Zhejiang University}
\email{simpson42@foxmail.com}
\author{Yi Cao}
\affiliation{Zhejiang University}
\email{gwater@zju.edu.cn}
\author{Congfu Xu}
\affiliation{Zhejiang University}
\email{xucongfu@cs.zju.edu.cn}
\authornote{Corresponding author.}
\settopmatter{printacmref=false} % Removes citation information below abstract
\acmConference{CIKM'18}{October 22-26, 2018}{Lingotto, Turin, Italy}
\setcopyright{rightsretained}
\begin{abstract}
\noindent Recommender systems are aimed at generating a personalized ranked list of items that an end user might be interested in. With the unprecedented success of deep learning in computer vision and speech recognition, recently it has been a hot topic to  bridge the gap between recommender systems and deep neural network. And deep learning methods have been shown to achieve state-of-the-art on many recommendation tasks. For example, a recent model, NeuMF, first projects users and items into some shared low-dimensional latent feature space, and then employs neural nets to model the interaction between the user and item latent features to obtain state-of-the-art performance on the recommendation tasks. NeuMF assumes that the non-interacted items are inherent negative and uses negative sampling to relax this assumption. 
% However, most of the existing methods only use neural network to incorporate side information to regularize latent user- or item-specific factors. Limited attention has been paid to explore full deep learning treatment for pure collaborative filtering setting.   the number of positive instances is equal to that of negative instances, which gives us hope to avoid the {\it Class Imbalance Problem}
In this paper, we examine an alternative approach which does not assume that the non-interacted items are necessarily negative, just that they are less preferred than interacted items. Specifically, we develop a new classification strategy based on the widely used pairwise ranking assumption. We combine our classification strategy with the recently proposed neural collaborative filtering framework, and propose a general collaborative ranking framework called Neural Network based Collaborative Ranking (NCR). We resort to a neural network architecture to model a user's pairwise preference between items, with the belief that neural network will effectively capture the latent structure of latent factors.  The experimental results on two real-world datasets show the superior performance of our models in comparison with several state-of-the-art approaches. 
% We also conduct extensive experiments to analyze the performance in different settings.
\end{abstract}

\keywords{recommender systems; neural networks; pairwise ranking}
\maketitle
\section{Introduction}
With the explosive growth of social media and e-commerce, we are living in the era of information explosion. Personalized recommendation is developed to alleviate the dilemma of information overload, and has become a core component of many popular e-commerce and social media services. Collaborative filtering (CF) \cite{hu2008collaborative,salakhutdinov2007probabilistic,Pan2008} is the most popular approach to personalized recommendation, which has been extensively studied in the past years. The two broad categories of CF are neighborhood-based approaches and model-based approaches. Neighborhood-based approaches, such as itemkNN \cite{Sarwar2001ICF}, first employ similarity metric to identify a set of similar items, and then generate top-$N$ recommended items based on those similar items. They can give explainable recommendations, but the relevance of their recommendation is lower in comparison with model-based methods. Model-based methods \cite{hu2008collaborative,Cremonesi2010PRA,salakhutdinov2007probabilistic}, especially latent factor models (LFM), map both users and items into a joint low-dimensional latent space. The prediction for a user's preference on an item is estimated by the inner product of the corresponding user and item latent vectors.

Implicit feedback refers to the scenarios where there are examples of items users prefer, but lack of examples of items they dislike, e.g., retweeting history in twitter, purchase history in e-commerce. LFM have achieved state-of-the-art on many recommendation tasks, however, traditional LFM methods suffer from the implicitness in implicit feedback. To address this issue, several variants of LFM has been proposed. For example, \citeauthor{hu2008collaborative} \cite{hu2008collaborative} proposed a Weighted Regularized Matrix Factorization (WRMF) method that weights observed and unobserved ratings differently and solves a regularized Least-Squares problem. \citeauthor{Rendle2009} \cite{Rendle2009} proposed Bayesian Personalized Ranking (BPR) that formulates $Top$-$N$ recommendation as a ranking problem and optimizes the Bayesian pairwise ranking criterion, which is the maximum a posteriori (MAP) estimation of users' pairwise preference between observed and unobserved items. 

In the literature \cite{He2017NCF,Hsieh2017CML}, it has been pointed out that the performance of LFM is hindered by using inner product as the user-item factor interaction function. For example, inner product does not guarantee the triangle inequality condition, as a result it is hard for the latent vectors of LFM to reliably capture the item-item or user-user similarity. To address this problem, \citeauthor{Hsieh2017CML} \cite{Hsieh2017CML} proposed to use metric learning\cite{kulis2013metric} to simultaneously capture users' preference and the user-user and item-item similarity. While \citeauthor{He2017NCF} \cite{He2017NCF} argued that inner product only captures linear interactions, proposing a general framework named NCF that employs neural networks to learn a function from data to effectively capture the nonlinear interaction between user factor and item factor. NCF uses the learned interaction function to replace inner product and gives promising results.
 % Recently, deep learning has been proven to be highly effective in many fields like computer vision \cite{he2016deep} and natural language processing tasks \cite{Collobert2008}. Motivated by this, many researchers have tried to bridge the gap between deep neural networks (DNNs) and recommender systems \cite{wang2015collaborative,Li2015,Kim2016CMF,He2016VBPR}. Nevertheless, these works mainly focus on leveraging deep learning models like autoencoders \cite{wang2015collaborative} or CNN to model side information (texts or images) to regularize latent user factors or latent item factors, and few efforts have been devoted to employing DNNs to perform recommenation tasks directly. And all methods above are based on matrix factorization, resorting to inner product to generate prediction. However, it has been pointed out that the simple choice of the user-item factor interaction function, inner product,  can hinder the prediction accuracy of LFM \cite{He2017NCF,Hsieh2017CML}. Inner product is a simple linear combination of the multiplication of latent factors, and may be insufficient to capture the complex relation between latent features.

% It is well known that neural network is capable of approximating any continuous function, which inspires us that we may employ DNNs that takes as input user factor and item factor to overcome the aforementioned problem. 
Nevertheless, we argue that the learning strategy of NCF might hinder its performance. NCF labels observed interactions as positive instances, and all unobserved interactions or some sampled unobserved interactions are labeled as negative instances. If all unobserved interactions are treated as negative instances, then it suffers from  two limitations: first, the negative class dominates the training data, which can degrade the predictive accuracy for the infrequent positive class (known as {\it Class Imbalance Problem}); second, treating non-interacted item as negative feedback does not conform to the facts that non-interacted item can also be interpreted as the user is not aware of it. Sampling some unobserved interactions as negative instances can alleviate the first problem, but it's still at the risk of introducing false negative examples.

% However, it simply labels observed item as positive instance and negative instance otherwise. We argue that NCF is hindered by two limitations: first, its classification strategy leads to the {\it Class Imbalance Problem}, i.e., the class distributions are highly imbalanced, which can degrade the predictive accuracy for the infrequent class; second, it treats uninteracted item as negative feedback, which does not conform to the facts that uninteracted item can also be interpreted as user is not aware of it.  
 
% Under our classification strategy, both positive and negative samples are pairwise,. One good characteristic of our classification strategy is that for each user the number of positive class is equal to that of negative class, which gives us hope to solve the {\it Class Imbalance Problem} in NCF

To tackle the problem above, in this paper, we develop a novel classification strategy for collaborative ranking. Based on the widely employed pairwise preference assumption that a user prefers observed items over all other unobserved items, we construct a positive preference set  and a negative preference set from rating data. Then, the elements in positive preference set are labeled as 1, and 0 for negative preference set, as seen in section 4.2. Our classification strategy has the following advantage: (1) The total number of positive examples are equal to the total number of negative examples, which gives us hope to solve the {\it Class Imbalance Problem}; (2) Under our classification strategy, negative instances do not assume non-interacted items to be negative feedback, just that they are less preferred than interacted items. Finally, we combine the proposed classification strategy with NCF to present a neural collaborative ranking framework. 

% And we focus on the implicit feedback scenarios, e.g., user purchases a product or likes a tweet, because implicit feedback is more prevalent and easier to collect. 

% To summarize, this work mainly makes the following contributions:
% \begin{itemize}
% 	\item We address the limitations of NCF, and develop a novel classification strategy to overcome it. 
% 	% To the best of our knowledge, our proposed NCR framework is the first to adopt the binary cross-entropy loss for pairwise ranking based recommendation.

%     \item We develop a novel and general recommendation algorithm based on neural networks and pairwise preference assumption.

% 	% \item We present a general framework named NCR for collaborative ranking based on neural network. To the best of our knowledge, NCR is the first to adopt the binary cross-entropy loss for pairwise ranking based recommendation.
	
% 	% \item We demonstrate that BPR-MF can be interpreted as a specialization of NCR. By selecting a particular scoring function for latent features, BPR-MF can be recovered under NCR on the fly. %The big difference lies in NCR can learn the non-linear interactions between latent features. 

% 	\item The experimental results on two real-world datasets show the superior performance of our proposed NCR in comparison with several state-of-the-art baselines. 

% \end{itemize}

\section{RELATED WORK}
\subsection{One-Class Collaborative Filtering}
When it comes to implicit feedback, we cannot simply treat the non-interacted items as negative examples, because the reason why a user doesn't interact with an item is ambiguous, e.g., the user may dislike it or the user is just unaware of it. Implicit feedback scenarios are also referred to as {\it one-class collaborative filtering ({\it OCCF})} problems \cite{Pan2008}. One crucial issue of {\it OCCF} is lack of negative feedback. Matrix Factorization (MF) is the most popular collaborative filtering technique. However, traditional MF approaches are incapable of handling the no negative feedback problem of {\it OCCF}. Because if the missing user-item interactions are treated as negative samples or just ignored, then the learner cannot generalize well.
%  MF derives a real-valued latent vector for each user and item and the user-item preference score is estimated by vector inner product.

 To tackle the above problem, several approaches have been proposed. According to how the missing data is used, existing methods to {\it OCCF} can be classified into two categories e.g., sampling based approaches \cite{He2017NCF,Pan2008,Rendle2009} and whole-data based approaches \cite{hu2008collaborative}. The former samples negative examples from the unobserved user-item interactions, while the latter includes all the unobserved user-item interactions as negative examples and uses a conditional weight to demote the influence of these ambiguous examples. We can also mainly categorize these approaches into point-wise methods and pairwise methods, according to how the relevance order is learned. Point-wise approaches generally regard user ratings as categorical labels or numerical values, and try to learn the relevance scores of missing data directly. While pairwise approaches try to capture the preference order between missing data. Pairwise approaches generally improve the ranking performance over point-wise approaches \cite{balakrishnan2012collaborative,lee2014local}.

\subsection{Deep Neural Networks}
There are many existing works trying to bridge the gap between deep neural networks (DNNs) and the task of collaborative filtering. A pioneering work along this direction is proposed by \cite{salakhutdinov2007restricted}, they adopted a variant of Restricted Boltzmann Machine, which is a two-layer undirected graphical model consisting of softmax visible units and binary hidden units, to perform the task of rating prediction. Hybrid collaborative filtering methods,  which combine deep learning with MF, has received much attention recently. These work mainly focuses on leveraging deep learning models like autoencoders \cite{wang2015collaborative,Li2015} or CNN to model side information (texts or images) to regularize latent user or item factors. Typical approaches include Collaborative Deep Learning (CDL) \cite{wang2015collaborative} and Convolutional Matrix Factorization (ConvMF) \cite{Kim2016CMF}. The former employed SDAE \cite{vincent2010stacked} to model texts, while the latter argued that bag-of-words model like SDAE has an inherent drawback, and proposed to use CNN to learn more effective latent features. In spite of their promising results, these approaches generally try to integrate deep learning with conventional recommender systems, no much attention has been paid to applying deep learning to develop pure collaborative filtering approaches to {\it OCCF}.
% For more works about deep learning methods on recommender systems, we direct readers to this survey \cite{zhang2017deep}. 
% proposed deep matrix factorization (deepMF) models, which first employ neural network with non-linear activation to map user and item rating vectors into a shared low-dimensional space, and then the relevance of user to each items is calculated by cosine similarity between the corresponding low-dimensional vectors. Their work is inspired by recent advance in web serach, where a network architecture named DSSM was proposed to map the query and documents into a common low-dimensional space, and to calculate the relevance of query and documents based on cosine similarity.
 
 Another line of work tries to use deep learning to make recommendation directly. For example, \citeauthor{cheng2016wide} \cite{cheng2016wide} proposed a context-aware recommendation method called Wide\&Deep, which first embedded features into latent space then used a multi-layer perceptron (MLP) on the concatenation latent vectors to learn the latent structure. The idea using MLP on the concatenation latent vectors latter was modified by \citeauthor{He2017NCF} \cite{hu2008collaborative}, they proposed a general framework for neural network based collaborative filtering (NCF). NCF takes advantage of the one-class nature of implicit feedback and casts {\it OCCF} as a binary classification problem. More recently, the attention mechanism has been introduced to the task of collaborative filtering. In \cite{chen2017attentive}, the proposed model ACF adopted item- and component-level attention to address the implicitness in users' interactions with multimedia content. ACF used two attention sub-networks to capture user's preference degree in item level and component level. Item-level attention was employed to score the item preferences, while the component-level attention was employed to capture interesting components in multimedia content. Again, it is worth highlighting that most of these work focuses on recommendation scenarios with rich feature, while no much attention has been paid to deep learning for pure collaborative filtering approaches to {\it OCCF}.

 % , which employs a multi-layer perceptron to learn the non-linear interactions between user and item latent features. 

 % NCF is a pointwise method, it takes advantage of the one-class nature of implicit feedback, labeling the observed items as 1 and unobserved items as 0. Then, unlike conventional pointwise performing regression with square loss, NCF minimizes the binary cross-entropy.  
%Among all these works, NCF is the closest to our work. The big different lies in: NCF is a pointwise method, while NCR is a pairwise method; NCF simply considers observed items as label 1 otherwise 0, while NCR considers user preference over a pair of observed and unobserved items as label 1 or 0; NCF 

\section{PRELIMINARIES}
Assume that we have a set of users $\mathcal{U}$ and a set of items $\mathcal{I}$, with $m=|\mathcal{U}|$ and $n=|\mathcal{I}|$ respectively. $R\in\{0,1\}^{m\times n}$ is the user-item rating matrix, where $R_{ui}$ indicates whether user $u$ rated item $i$ or not. We denote by $\mathcal{I}_u^+$ the set of the items rated by user $u$. Matrices $U\in \mathbb{R}^{k\times m}$, $V\in \mathbb{R}^{k\times n}$ are the latent representations of users and items respectively, $U_i$ denotes the $i$-th column of $U$, and $V_i$ likewise. The goal in {\it OCCF} is to obtain a predicted ranking over items.

\subsection{Latent Factor Models}
The basic idea of latent factor models is to transform both users and items into some shared low-dimensional latent feature space. Matrix factorization is the most popular technique to derive latent factor models. Formally, let denote $W$ as some weighting matrix, the objective of MF is to minimize the following regularized squared loss:
\begin{equation}
\min_{U,V}\sum_{i\in \mathcal{U}, j\in \mathcal{I}}W_{ij}\cdot(R_{ij}-U_i^TV_j)^2+\lambda(||U||^2+||V||^2),
\end{equation}
where $\lambda$ is the regularization hyperparameter. One classical MF approaches is Singular Value Decomposition (SVD). In SVD, $W$ is conditioned on whether a user has interacted with an item or not, e.g., $W_{ij}=\text{\uppercase\expandafter{\romannumeral2}}(R_{ij}>0)$, where \text{\uppercase\expandafter{\romannumeral2}($\cdot$)} denotes the indicator function that returns 1 if the statement is true and 0 otherwise. This weighting scheme is inappropriate in implicit scenarios, as it will lead to trivial but useless solutions (e.g., all the miss entries of $R$ is predicted as 1). An alternative  approach is to use some weighting scheme to give larger weight to observed ratings meanwhile small but non-zero weight to unobserved ratings, which leads to the WRMF method \cite{hu2008collaborative}.
% One classical MF approches is Singular Value Decomposition (SVD)
\subsection{Bayesian Personalized Ranking (BPR)}
An alternate strategy to address the implicitness in {\it OCCF} is BPR. Instead of predicting the relevance scores directly, BPR models a user's preference over two items, where one of the item is observed and the other is not. BPR is a well-known pairwise ranking optimization framework, which assumes that the known positive preferences over observed items are ranked higher than all the other unknown preferences over unobserved items. Let $D_S$ denote the triplets of the form $(u$, $i$, $j)$, $u$ is a user, $i$ is an observed item, $j$ has not been observed yet:
\begin{equation}
D_S=\{u\in \mathcal{U}, i\in \mathcal{I}_u^+, j\in\mathcal{I}\wedge j\notin\mathcal{I}_u^+\}
\end{equation}
BPR optimizes a loss over a {\it(user, item, item)} triplet, the following optimization criterion is used for personalized ranking (BPR-OPT):
\begin{equation}
\sum_{(u,i,j)\in D_S }-ln\sigma(\hat{x}_{uij})+\lambda_\Theta||\Theta||^2,
\end{equation}
where $\sigma(x)=\frac{1}{(1+e^{-x})}$ is the sigmoid function, $\Theta$ is the parameter vectors, $\lambda_{\Theta}$ is the regularization hyperparameter. In particular, BPR-MF is obtained when $\hat{x}_{uij}$ is predicted by matrix factorization:
\begin{equation} \hat{x}_{uij}=U_u^TV_i-U_u^TV_j. \end{equation}
% \subsection{The limitation of Latent Factor Models}

\section{Proposed Method}
In this section, we will introduce our Neural Collaborative Ranking (NCR) model in detail. We first describe our neural network based pairwise ranking model, elaborating how to learn NCR with a probabilistic model that emphasizes user preference over a pair of observed and unobserved items. We then show the relations between our model and BPR-MF, and develop a shallow model using linear interactions between latent vectors. 
Next, a deep instantiation of NCR using multi-layer perceptron to model latent features is proposed to investigate deep neural networks for collaborative ranking. MLP endows our model with a high level of nonlinearities. Finally, a new pairwise ranking model unifying the strengths of linear and nonlinear interactions for modeling latent features is presented. 
% \subsection{Background and Notation}
% Assume that we have a set of users $\mathcal{U}$ and a set of items $\mathcal{I}$, with $m=|\mathcal{U}|$ and $n=|\mathcal{I}|$ respectively. $R\in\{0,1\}^{m\times n}$ is the user-item rating matrix, where $R_{ui}$ indicates whether user $u$ rated item $i$ or not. We denote by $\mathcal{I}_u^+$ the set of the items rated by user $u$. Matrices $U\in \mathbb{R}^{k\times m}$, $V\in \mathbb{R}^{k\times n}$ are the latent representations of users and items respectively, $U_i$ denotes the $i$-th cloumn of $U$, and $V_i$ likewise. BPR is a pairwise ranking optimization framework, which assumes that the known positive preferences over observed items are at least as high as the unknown preferences over unobserved items. Let $D_S$ denote the triples of the form $(u,i,j)$, $u$ is a user, $i$ is an observed item, $j$ has not been observed yet:
% \begin{equation}
% D_S=\{u\in \mathcal{U}, i\in \mathcal{I}_u^+, j\in\mathcal{I}\wedge j\notin\mathcal{I}_u^+\}
% \end{equation}
% BPR optimizes a loss over {\it(user,item,item)} triples, the following optimization criterion is used for personalized ranking (BPR-OPT):
% \begin{equation}
% \sum_{(u,i,j)\in D_S }ln\sigma(\hat{y}_{uij})-\lambda_\Theta||\Theta||^2,
% \end{equation}
% where $\sigma(x)=\text{log}(1+e^{-x})$ is the logistic loss, $\Theta$ is the parameter vector, $\lambda_{\Theta}$ is the regularization hyperparameter. In particular, BPR-MF is obtained when $\hat{x}_{uij}$ is predicted by matrix factorization:
% \begin{equation} \hat{y}_{uij}=U_u^TV_i-U_u^TV_j. \end{equation}

\subsection{General Framework}
We now elaborate NCR, our proposed general framework for collaborative ranking based on neural network. In order to obtain a full neural treatment of collaborative ranking, following Wide\&Deep, we adopt feed-forward neural networks to model a {\it (user, item, item)} triplet interaction $\hat{y}_{uij}$, as shown in Figure 1. Our model consists of three layers, the bottom embedding layer, the middle hidden layers and the output prediction layer. Hereinafter, we elaborate the neural network architecture layer by layer.
\begin{figure}[ht]
\centering
\includegraphics[scale=0.5]{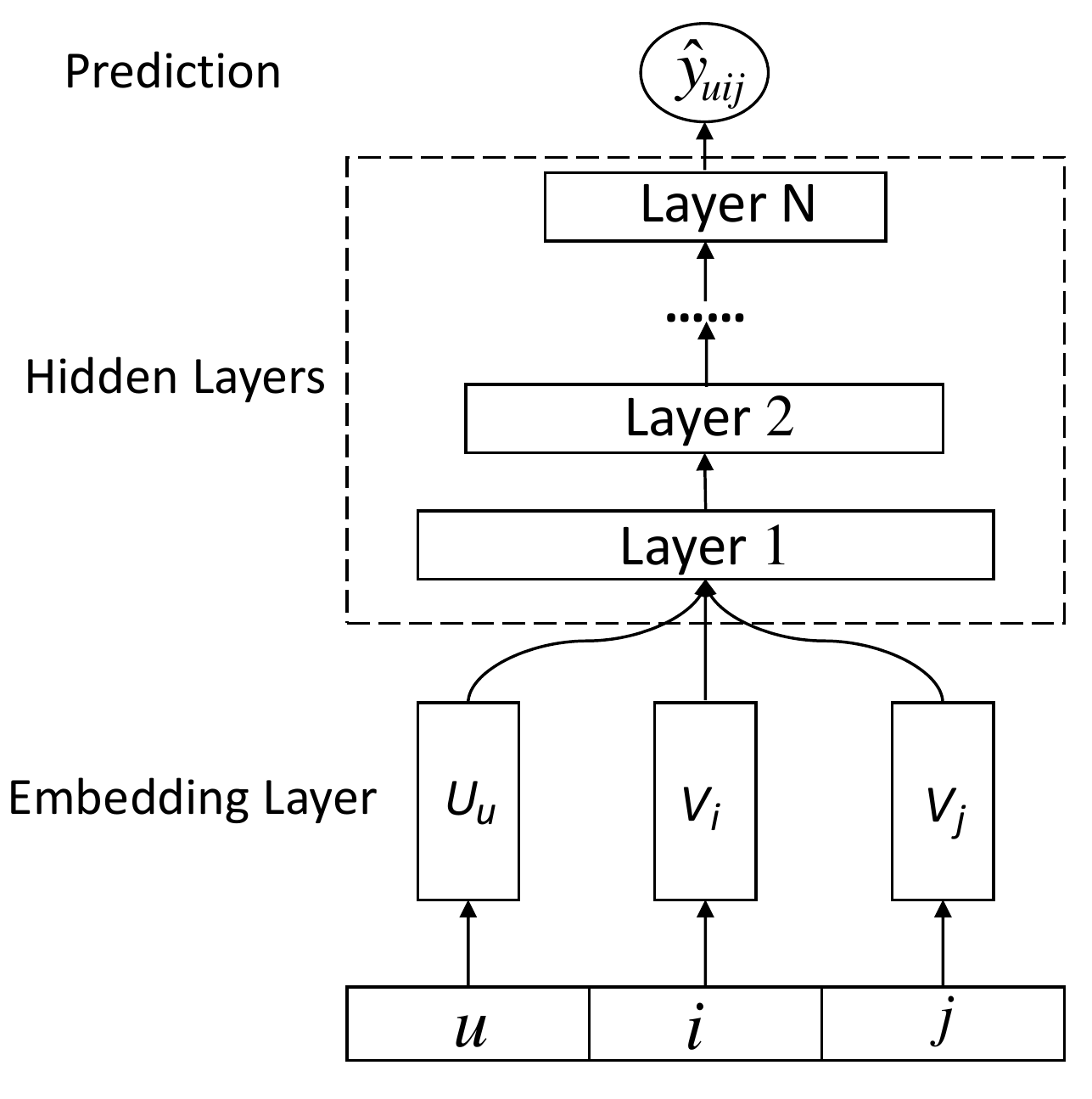}
\caption{Network Architecture for Neural Collaborative Ranking}
\label{ncr1}
\end{figure}

\textbf{Embedding Layer}. The goal of the embedding layer is to transform both users and items into some shared low-dimensional latent feature space. After embedding, we acquire a dense vector representation for each user and item. This shares the same spirit of LFM mentioned before. Formally, let $(u, i, j)$ be an input triplet, we use embedding table lookup to obtain three embedding vectors $U_u, V_i$ and $V_j$, respectively. The embedding layer can be easily extended to cover a wide range of auxiliary information, such as topic information \cite{wang2011collaborative} and multimedia content \cite{chen2017attentive}. Since in this work we only focus on the pure collaborative ranking setting, we do not take any side information into account. 

\textbf{Hidden Layers}. The hidden layers are a stack of fully connected layers built above the embedding layer. The obtained dense vectors from embedding layer are concatenated together, resulting in a dense vector jointly encoding user preference and item attribute. Then the concatenated vector is fed into the hidden layers. Hidden layers are the key to endow our model with the capacity to learn highly nonlinear interactions between latent features. In particular, the size of the last hidden layer determines the model's capability, so we term it as {\it predictive factors}. 

Let $N$ be the number of hidden layers, the concatenated vector are propagated forward layer by layer, so we
can formulate the interaction function $f$ as follow:

\begin{equation}
f(U_u,V_i,V_j)=f_N(\cdots f_2(f_1(U_u,V_i,V_j))), 
\end{equation}
where $f_l (l=1, 2, \cdots, N)$ denotes the mapping function for the $l$-th hidden layer.

% $f_{out}$ and g(U_u,V_i,V_j|U,V,\Theta_g), g(U_u,V_i,V_j)&=,\\
\textbf{Prediction Layer}.
 The prediction layer maps previous layers' output to the prediction score $\hat{y}_{uij}$. $\hat{y}_{uij}$ expresses the extent user $u$ prefers item $i$ to item $j$. The prediction score given by NCR can be formulated as follows:
\begin{equation}
\hat{y}_{uij}=f_{out}(f_N(\cdots f_2(f_1(U_u,V_i,V_j)))),
\end{equation}
where $f_{out}$ denotes the the mapping function for the output layer. In our case, it's the sigmoid function.

In this paper we choose to use a set of unified hidden layers to model the latent structure of a {\it (user, observed item, unobserved item)} triplet. Another choice is to employ two sets of hidden layers to model {\it (user, observed item)} and {\it (user, unobserved item)} pairs, respectively. Then the prediction score is formulated as follows:
\begin{equation}
g(U_u,V_i)=f_{out}(f_N(\cdots f_2(f_1(U_u,V_i))))
\end{equation}
\begin{equation}
g(U_u,V_j)=f_{out}(f_N(\cdots f_2(f_1(U_u,V_j))))
\end{equation}
\begin{equation}
\hat{y}_{uij}=g(U_u,V_i)-g(U_u,V_j)
\end{equation}
 We consider the former being more feasible, the intuition behind is that it also takes the nonlinear interactions between items into account. For the latter, we leave it as a future work.

\subsection{Model Learning}\quad To learn the parameters of our models, it's straightforward to adopt the widely used logistic ranking loss as the loss function:
\\
\begin{equation}
\mathcal{L}_{\text{log}}=-\sum_{(u,i,j)\in D_S}\text{log}(1+e^{-\hat{y}_{uij}})
\end{equation}
% or the squared loss \cite{Yao2015CTRank}:
% \begin{equation}
% \mathcal{L}_{\text{sq}}=-\sum_{(u,i,j)\in D_S}(1-e^{-g(U_u,V_i,V_j)})^2
% \end{equation}
However, logistic loss may suffer from vanishing gradients for correctly ranked pairs \cite{Rendle:2014:IPL:2556195.2556248}. Besides, prior work \cite{burges2005learning,He2017NCF} shows that binary cross-entropy loss is a good choice for neural network based ranking, so we adopt the binary cross-entropy loss for our model. Hereinafter, we demonstrate that under our classification strategy, it is nature to formulate the binary cross-entropy for learning with NCR. And we elaborate how to construct the positive instances and negative instances for training.
\\\\
\indent\textbf{Classification Strategy}. The use of log ranking loss is based on the assumption that a user prefers observed items to unobserved items:
\begin{equation}
\forall (u,i,j)\in D_S,i>_uj,
\end{equation}
where $>_u\subset \mathcal{I}^2$ is the personalized total ranking \cite{Rendle2009}. We call $D_S$ positive preference set. Similarly, we can construct a negative preference set:
\begin{equation}
\bar{D}_S=\{ (u,j,i): (u,i,j)\in D_S \}
\end{equation} 
For all triples in set $D_S\cup \bar{D}_S$, we define the following indicator function:
\begin{equation}
y_{uij}= \begin{cases} 1, \text{if } (u,i,j)\in D_S\\
                       0, \text{if } (u,i,j)\in \bar{D}_S 
          \end{cases}
\end{equation} 

Then we view the value of $y_{uij}$ as a label 1 for $(u,i,j)\in D_S$; for all triplets in $\hat{D}_S$, we view the value of $y_{uij}$ as a label 0. It's obvious that the size of $D_S$ is equal to $\bar{D}_S$. Thus, we successfully avoid the {\it Class Imbalance Problem} in NCF. We constrain the prediction score $\hat{y}_{uij}$ in the range of [0,1], and interpret it as how likely the triple belongs to $D_S$. With the above settings, the likelihood function is defined as follows:
\begin{equation}
p(D_S,\bar{D}_S|U,V,\Theta_g)=\prod_{(u,i,j)\in D_S}\hat{y}_{uij}\prod_{(u,i,j)\in \bar{D}_S}(1-\hat{y}_{uij})
\end{equation}
Take the negative logarithm of the likelihood function, we endow our NCR with the binary cross-entropy loss
\begin{equation}
\begin{split}
\mathcal{L}=&-\sum_{(u,i,j)\in D_S}\text{log}\hat{y}_{uij}-\sum_{(u,i,j)\in \bar{D}_S}\text{log}(1-\hat{y}_{uij})\\
=&-\sum_{(u,i,j)\in D_S\cup \bar{D}_S}y_{uij}\text{log}\hat{y}_{uij}+(1-y_{uij})\text{log}(1-\hat{y}_{uij})
\end{split}
\end{equation}
%&-\sum_{(u,i,j)\in D_S}\text{log}\hat{y}_{uij}-\sum_{(u,i,j)\in \bar{D}_S}\text{log}(1-\hat{y}_{uij})

\textbf{Discussion}. The classification strategy above is based on the widely used pairwise preference assumption. By labeling a triplet $(u, i, j)\in D_S$ as a positive instance, our model learns to rank an item $i$ (known preference) higher than an item $j$ (unknown preference). Likewise, By labeling a triplet $(u, i, j)\in \bar{D}_S$ as a negative instance, our model learns to rank an item $i$ (unknown preference) lower than an item $j$ (known preference). As a result, both positive and negative instance contribute to the pairwise ranking process. Since we employ the binary cross-entropy loss, the negative instance is necessary, which is different from the log ranking loss.

% To the best of our knowledge, this is the first algorithm that adopts binary cross-entropy loss for pairwise ranking based recommendation.

By utilizing a probabilistic treatment for NCR, we address pairwise ranking based recommendation as a binary classification problem.  At training stage, we uniformly sample positive and negative instances from $D_S$ and $\bar{D}_S$ respectively. In practice, we iteratively update the parameters until the loss does not decrease (by 0.1\%) or the maximum iteration limit is reached. In latter section, we also conduct experiments to study the influence of negative samples on the results. 

\subsection{Relations to Other Methods}
\subsubsection{Relations to Bayesian Personalized Ranking} BPR-MF can be seen as a special case of NCR without hidden layers. In what follows, we concretely show that if we choose specific interaction function, output function and edge weight, our NCR model will degenerate into BPR-MF. As BPR-MF is the most popular method for pairwise ranking based recommendation, the fact that BPR-MF can be explained as a special case of NCR reveals that it is trivial for NCR to accommodate a wide range of pairwise ranking approaches.

To recover BPR-MF, we set the interaction function of latent vectors as
\begin{equation}
f_1(U_u,V_i,V_j)=\begin{bmatrix}
U_u\odot V_i\\
-U_u\odot V_j
\end{bmatrix},
\end{equation}
where $\odot$ denotes the element-wise product of vectors. This vector is then project to the output layer
\begin{equation}
\hat{y}_{uij}=a_{out}(\textbf{w}^T\begin{bmatrix}
U_u\odot V_i\\
-U_u\odot V_j
\end{bmatrix}),
\end{equation}
where $a_{out}$ and $\textbf{w}$ is the activation function and the edge weight of the output layer, respectively. Then we define $a_{out}$ as  
\begin{equation}
a_{out}(x)=-\text{log}(1+e^{-x}),
\end{equation}
and let $\textbf{w}$ be a vector with all elements equal to 1. In this way, we can obtain the BPR-MF model. 

In this work, we use the sigmoid function as $a_{out}$, since we want to constrain the output in the range of [0,1]. For the edge weight $\textbf{w}$, instead of constraining it to be a vector of 1, we learn it from data, which allows $\textbf{w}$ to vary the importance of latent dimensions. We term the degenerated model as NBPR, short for neural Bayesian personalized ranking.

\subsubsection{Relations to RankNet} RankNet \cite{burges2005learning} is a well-known pairwise ranking method for information retrieval tasks.  Our proposed model NCR shares some similarities with RankNet, e.g., both models are based on neural network and adopt binary cross-entropy loss. Nevertheless, RankNet was originally proposed for information retrieval tasks with dense features, which might not be directly applied in {\it OCCF} setting with no context information. We also mainly address the following difference: RankNet is a "point-wise" model endowed with pairwise ranking policy, while NCR itself is a "pairwise" model using pairwise ranking policy. By "point-wise", we mean both the input and output of RankNet are point-wise, i.e., it takes as input one training sample at a time, and the output is also a predicted score for a single sample. Pairwise ranking in RankNet is conducted by minimizing the loss function of two consecutive training samples. While NCR is inherently "pairwise" as it takes as input a pair of items at a time, the output score is also a predicted preference over two items. 

\subsection{Predictive Rule}
As for how to make recommendation, we cannot sort the output scores directly to obtain the top-$N$ ranked items, because an output score is associated with a pair of items rather than a single item. To get rid of this bad situation, we provide a heuristic approach. Before making recommendation, let us discuss the consistency requirement in pairwise ranking.

Ideally, given a user $u$, three items $i, j$ and $k$, if our model asserts $i>_u j$ and $j>_u k$, we also want it to assert $j>_u k$. Otherwise it would be hard to rank the three items correctly. Note that the consistency requirement in our case is different from RankNet, as we cannot calculate the combining probabilities for $i$ and $k$. In what follows, we show that under certain conditions, our model indeed can meet the consistency requirement. Recall that in section 4.3.1, we present a NCR model NBPR. We rewrite the edge weight \textbf{w} in NBPR as 
$$\mathbf{w}=\begin{bmatrix}
\mathbf{w}_1\\
\mathbf{w}_2
\end{bmatrix},$$
where the dimensionality of $\mathbf{w}_1$ is equal to $\mathbf{w}_2$. Then, we have the following predicted scores:
\begin{equation}
\begin{split}
\hat{y}_{uij}&=a_{out}(\mathbf{w}_1^T\cdot U_u\odot V_i-\mathbf{w}_2^T\cdot U_u\odot V_j),\\
\hat{y}_{uji}&=a_{out}(\mathbf{w}_1^T\cdot U_u\odot V_j-\mathbf{w}_2^T\cdot U_u\odot V_i),\\
\end{split}
\end{equation}
where $a_{out}$ is given by Equation 18 and is a monotonically increasing function. We have similar results for $\hat{y}_{ujk}, \hat{y}_{ukj}, \hat{y}_{uik}$ and $\hat{y}_{uki}$. The first problem is how to rank two items $i$ and $j$. Intuitively, $\hat{y}_{uij}$ indicates the probability that $u$ likes $i$ more than $j$, while $\hat{y}_{uij}$ indicates the probability that $u$ likes $j$ more than $i$. So we can infer that if $\hat{y}_{uij}>\hat{y}_{uji}$, then the user $u$ will like $i$ more than $j$. As a result, we give the following predictive rule:
{\it If} $\hat{y}_{uij}>\hat{y}_{uji}$, \text{\it then} $i>_uj$; {\it otherwise} $j>_ui$. According to the predictive rule above, if $i>_uj$ and $j>_uk$, we have
\begin{equation}
\mathbf{w}_1^T\cdot U_u\odot V_i-\mathbf{w}_2^T\cdot U_u\odot V_j>\mathbf{w}_1^T\cdot U_u\odot V_j-\mathbf{w}_2^T\cdot U_u\odot V_i,
\end{equation}
\begin{equation}
\mathbf{w}_1^T\cdot U_u\odot V_j-\mathbf{w}_2^T\cdot U_u\odot V_k>\mathbf{w}_1^T\cdot U_u\odot V_k-\mathbf{w}_2^T\cdot U_u\odot V_j.
\end{equation}
\noindent Add the above two equations and eliminate duplicates, we have
\begin{equation}
\mathbf{w}_1^T\cdot U_u\odot V_i-\mathbf{w}_2^T\cdot U_u\odot V_k>\mathbf{w}_1^T\cdot U_u\odot V_k-\mathbf{w}_2^T\cdot U_u\odot V_i.
\end{equation}
\noindent In other words, we have $\hat{y}_{uik}>\hat{y}_{uki}$, i.e., $i>_uk$. In consequence, the consistency requirement is met. For NCR model with hidden layers, if we use two sets of hidden layers to model user's interactions between observed item and unobserved item, respectively, the above conclusion can also hold. However, if we decide to use a unified hidden layers to model the interactions, we have no idea whether the predictive rule above can meet the consistency requirement or not, so we call it a "heuristic" approach, and our experimental results show it works well.

Based on the analysis above, we propose a simple algorithm to find the top-$K$ ranked items, as shown in Algorithm 1. To rank two items $i$ and $j$, we need to compare the predicted score $\hat{y}_{uij}$ and $\hat{y}_{uji}$. Algorithm 1 scans the candidate item set $K$ pass, each pass choosing a most preferred item. 

\begin{algorithm}
\caption{NCR Recommendation}
 \SetKwInOut{Input}{Input} \SetKwInOut{Output}{Output}
 \Input{A user $u$; A set of unobserved items $\mathcal{I}_u^-=\{I_1,\cdots,I_m\}$; number of recommendation $K$; ranked list $\mathcal{P}_{uk}\leftarrow\phi$}
 \Output{top-$K$ items $\mathcal{P}_{uk}$}
 \For{$p\leftarrow 1$ \KwTo $K$}{
 $\mathcal{I}_u^-$=$\mathcal{I}_u^-$ - $\mathcal{P}_{uk}$\\
 $M$=length of $\mathcal{I}_u^-$\\
 max=$\mathcal{I}_u^-[1]$\\
 \For{$i\leftarrow 2$ \KwTo $M$}{
 j=$\mathcal{I}_u^-[i]$\\
 \If{$\hat{y}_{(u, j, {\rm max})}$>$\hat{y}_{(u, {\rm max}, j)}$}{max=j}
}
Append max to the end of $\mathcal{P}_{uk}$
}
% $\mathcal{P}_{uk}$=$\mathcal{P}_{u}$[1$\cdots K$]
\end{algorithm}

\subsection{Deep Neural Collaborative Ranking}
In order to make full use of DNNs' capacity, in this section we investigate how to go deep with NCR. In the embedding layer of NCR, there are embeddings for user and item, respectively. For every triplet in the training set, we have three latent vectors. Intuitively, we can concatenate these latent vectors together. This is a widely used technique in many existing deep learning work \cite{He2017NCF,NIPS2012_4683,NIPS2012_4683}. However, simply concatenating latent features is insufficient to capture the user-item latent structures, because it dose not take any interactions between latent dimensions into consideration. To address this problem, following Wide\&Deep, we add hidden layers on the concatenated vector. More precisely,  we employ a standard MLP to to capture the user-item latent structures. MLP's multi-layer nature enable it to learn a variety of levels of user-item latent structures, especially the nonlinear interactions between latent features. In comparison with NBPR using a fixed element-wise product between user and item latent vectors, MLP is more flexible when dealing with the concatenated vector. Proceeding on this track, we define a deep neural collaborative ranking (DNCR) model under NCR framework as 
\begin{equation}
\begin{split}
\textbf{x}_1=f_1(U_u,&V_i,V_j)=\begin{bmatrix}
U_u\\
V_i\\
-V_j
\end{bmatrix},
\\\textbf{x}_2=f_2(\textbf{x}_1)=&a_2(\textbf{W}_2^T\textbf{x}_1+\textbf{b}_2),
\\\cdots&\cdots
\\\textbf{x}_N=f_N(\textbf{x}_{N-1})=&a_2(\textbf{W}_N^T\textbf{x}_{N-1}+\textbf{b}_N),
\\\hat{y}_{uij}=\sigma(&\textbf{w}^Tf_N(\textbf{x}_{N-1})),
\end{split}
\end{equation}
where $\textbf{W}_*$ and $\textbf{b}_*$ denote the weights and biases respectively, $a_*$ is the activation function. We use tanh as activation functions of hidden layers. As for the network architecture,  we follow the popular tower pattern (e.g. \cite{cheng2016wide,He2017NCF}). The width (number of neurons) of a layer is decreased with its height (the number of layers below). More precisely, we first set the number of neurons in the bottom layer, then half the layer width for each successive higher layer. In this sense, higher layers can obtain more abstractive features.

\begin{figure}[ht]
\centering
\includegraphics[scale=0.45]{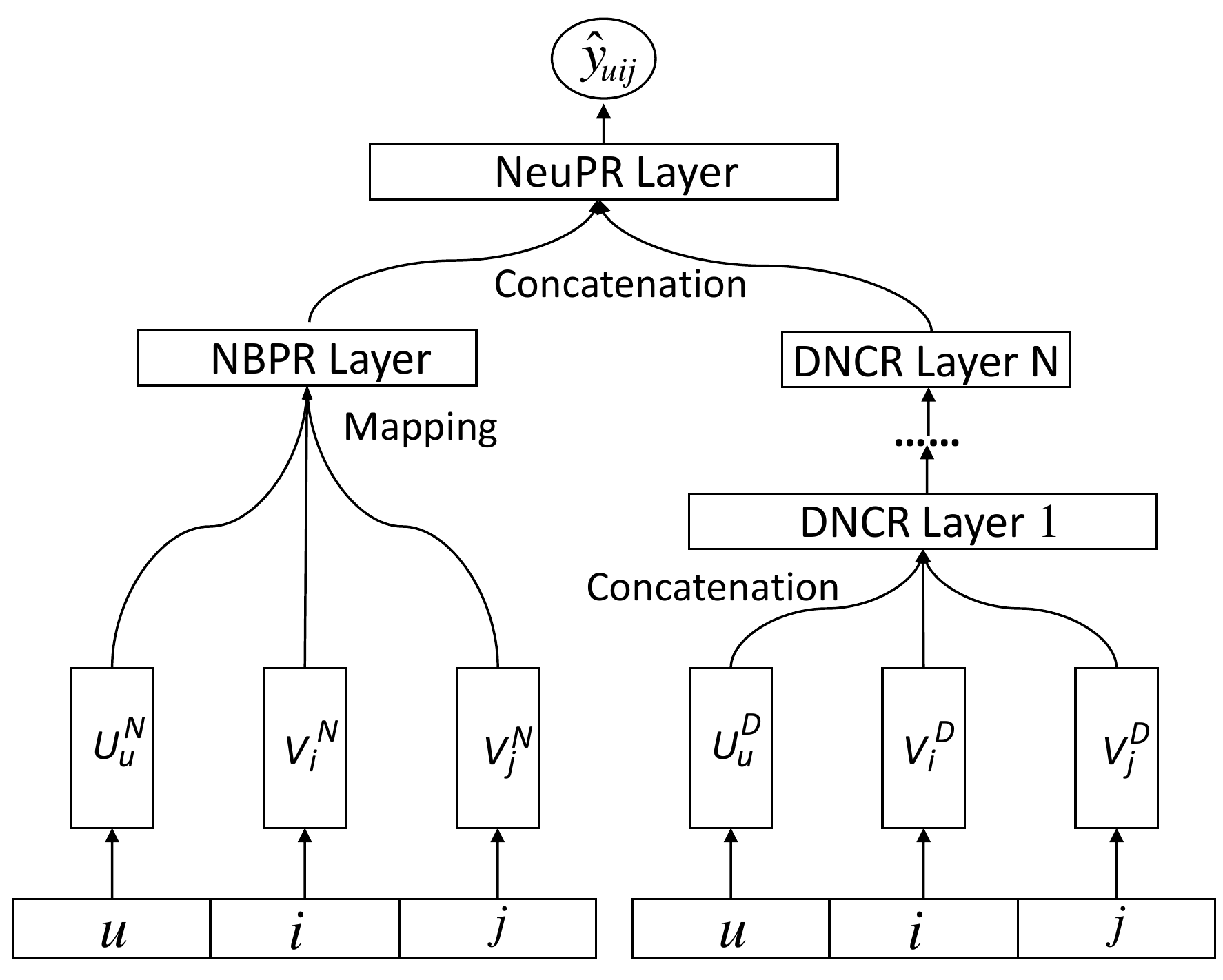}
\caption{Neural pairwise ranking model}
\label{ncr2}
\end{figure}

\subsection{Neural Personalized Ranking}
In this section, we develop a model to combine NBPR and DNCR. 
% Both NBPR and DNCR are instantiations of the NCR framework, by fusing them together, we may enjoy of advantage of both models simultaneously. 
The intuition behind is twofold: first, NBPR is a shallow model with limiting capacity, while DNCR is a deep model at the risk of overfitting, fusing them together can increase model's capacity at the same time prevent overfitting; second, NBPR applies a linear mapping to model the interactions of latent user and item vectors, DNCR applies a nonlinear kernel to model the latent structures of features, by fusing them together, we can obtain a model enjoying the advantage of linearity and nonlinearity simultaneously.  As for how to fuse them together, a trivial solution is to let NBPR and DNCR share the same embedding layer, and then fuse them together by combining the outputs of their learned interaction functions
\begin{equation}
\hat{y}_{uij}=\sigma(\textbf{w}^Ta(\begin{bmatrix}
U_u\odot V_i\\
U_u\odot -V_j
\end{bmatrix}+\textbf{W}^T_N\textbf{x}_{N-1}+\textbf{b}_N)),
\end{equation}
where $\textbf{x}_{N-1}$ is the $(N-1)$-th layer's output of MLP (Equation. 23). This solution is similar to Neural Tensor Network (NTN) \cite{socher2013reasoning}. However, due to the different learning process of the two models, their optimal embedding dimensionality and weights might be very different. Thus, constraining the two models to share the same embedding is not flexible enough, and may degrade the prediction performance. With this in mind, we propose to allow NBPR and DNCR to learn separate embeddings, and then fuse the two models by concatenating their last hidden layer, as shown in Figure 2. We formulize this solution as
\begin{equation}
\begin{split}
&f^{N}=\begin{bmatrix}
U_u^N\\
U_u^N
\end{bmatrix}\odot
\begin{bmatrix}
 V_i^N\\
 -V_j^N
\end{bmatrix},\\
&f^{D}=a_N(\textbf{W}^T_N(a_{N-1}(\dots a_2(\textbf{W}_2^T\begin{bmatrix}
U_u^D\\
V_i^D\\
-V_j^D
\end{bmatrix}+\textbf{b}_2)\dots))+\textbf{b}_N),\\
&\hat{y}_{uij}=\sigma(\textbf{w}^T\begin{bmatrix}
f^{N}\\
f^{D}
\end{bmatrix}),
\end{split}
\end{equation}
where $U^N, V^N$ denotes the user and item embeddings for NBPR part, respectively; and $U^D, V^D$ is similarly defined. We name this model NeuPR, short for neural personalized ranking. As we have discussed before, NeuPR enjoys the linearity of BPR and nonlinearity of MLP at the same time, thus may be able to yield better results than NBPR and DNCR. 

\subsection{Pre-training}
For NeuPR, randomly initialized weights and embeddings do not pass any information, thus it is hard for the output layers to capture meaningful features. As a result, the neural network cannot be trained effectively.
% the objective function is non-convexity, so it is hard to find the global optimum and easy to fall into local optima. 
On the other hand, improper initialization may lead NeuPR to being trapped in local optimum at an early stage, in consequence the convergence and performance suffer. To alleviate the problems above, it is intuitive to first train NBPR and DNCR with random initializations until convergence or the maximum iteration limit, then initialize NeuPR' NBPR part and DNCR part with the pre-trained models of NBPR and DNCR, respectively. The only modification is the edge weight, like prior work \cite{He2017NCF}, we concatenate the edge weights of the two pre-trained models with
\begin{equation}
\textbf{w}\leftarrow \begin{bmatrix}
\alpha \textbf{w}^N\\
(1-\alpha)\textbf{w}^D
\end{bmatrix},
\end{equation}
where $\textbf{w}^N, \textbf{w}^D$ denote the edge weight vector $\textbf{w}$ of NBPR and DNCR, respectively; and $\alpha$ is a hyperparameter which balances the trade-off between the two pre-trained models. 
% If $\alpha=0$, NeuPR is only pre-trained with the DNCR part, and if $\alpha=1$, NeuPR is only pre-trained with the NBPR part. Under these two extreme situation, the effect of pre-training declines.

\begin{table*}[!ht]
\centering
\begin{tabular}{c|cc|cccc|ccc|c}
\hline
\hline
number of &\multicolumn{1}{r}{\multirow{2}{*}{Datasets}} & \multirow{2}{*}{Metrics} & \multicolumn{4}{c|}{Methods} & \multicolumn{3}{c|}{NCR} & Improvement of\\ \cline{4-10}
 predictive factors               & &           &PopRank   & BPR  & eALS  &  NeuMF  & NBPR      & DNCR   & NeuPR       &  NeuPR vs. NeuMF           \\     \hline
\multirow{4}{*}{8} &\multirow{2}{*}{ML1m}& HR   & 0.4227  & 0.5096& 0.4861& 0.5480  & 0.5402   & \bf{0.5664} &  0.5661       & 3.30\%       \\
                                        && NDCG & 0.1815  & 0.2724& 0.2503& 0.2921  & 0.2854   & \bf{0.3037} &  0.2997       & 2.60\%          \\
\cline{2-11}

            & \multirow{2}{*}{Amusic}     & HR & 0.2710    & 0.2752&0.3263& 0.3476 &  0.3554   & \bf{0.3654}  & 0.3645       & 4.86\%        \\
                                        && NDCG& 0.1222    & 0.1586&0.1819& 0.2048 &  0.2062   & \bf{0.2207}  & 0.2124       & 3.71\%       \\ \hline

\multirow{4}{*}{16} &\multirow{2}{*}{ML1m} & HR& 0.454    &  0.5234&0.5156& 0.5515 &  0.5492   & 0.5672       & \bf{0.5692}  & 3.21\%         \\
                                       && NDCG &  0.254   &  0.2756&0.2691& 0.2989 &  0.2935   & 0.3076       & \bf{0.3124}  & 4.52\%         \\   
\cline{2-11}

       &  \multirow{2}{*}{Amusic}  &      HR&  0.229      & 0.2821&0.3247 & 0.3531 &  0.3643   & \bf{0.3862}  & 0.3707       & 4.98\%         \\
                                     & & NDCG  & 0.126    & 0.1623&0.1841 & 0.2016 &  0.2115   & \bf{0.2228}  & 0.2089       & 3.06\%          \\ 
\hline 

\multirow{4}{*}{24} &\multirow{2}{*}{ML1m}& HR & 0.454    & 0.5419& 0.5227& 0.5495  &   0.5478 &0.5690        &\bf{0.5724}   & 4.17\%   \\
                                        && NDCG& 0.254    & 0.2876& 0.2759& 0.2960  &   0.3002 &0.3082        &\bf{0.3104}   & 4.86\%   \\ 
      \cline{2-11}

          & \multirow{2}{*}{Amusic}        & HR& 0.229    & 0.2860& 0.3028 & 0.3505   & 0.3667  &\bf{0.3853}   & 0.3736       & 6.59\%    \\
                                        && NDCG& 0.126    & 0.1659& 0.1729 & 0.2022   & 0.2139  &\bf{0.2229}   & 0.2166       & 7.12\%    \\ \hline

\multirow{4}{*}{32} &\multirow{2}{*}{ML1m}& HR & 0.454    & 0.5513& 0.5344 & 0.5493  &  0.5435  & 0.5652      &\bf{0.5740}    & 4.50\%       \\
                                        && NDCG& 0.254    & 0.2895& 0.2833 & 0.2912  &  0.2973  & 0.3089      &\bf{0.3096}    & 4.14\%      \\ 
   \cline{2-11}

          & \multirow{2}{*}{Amusic       } & HR& 0.229    & 0.2958&0.2995  & 0.3441   &  0.3676   &\bf{0.3941} & 0.3758        & 9.21\%      \\
                                        && NDCG& 0.126    & 0.1738&0.1727  & 0.2006   &  0.2165   &\bf{0.2264} & 0.2185        & 8.92\%    \\ \hline

\multirow{4}{*}{64} &\multirow{2}{*}{ML1m}& HR & 0.454    & 0.5478& 0.5386 & 0.5400    &  0.5258  & 0.5753     &\bf{0.5801}    & 7.42\%       \\
                                        && NDCG& 0.254    & 0.2903& 0.2891 & 0.2987    &  0.2885  & 0.3116     &\bf{0.3158}    & 5.72\%      \\ 
  \cline{2-11}

          & \multirow{2}{*}{Amusic     } & HR  & 0.229   & 0.2920& 0.2961  & 0.3564    &  0.3793  &\bf{0.3904} & 0.3826        & 7.35\%     \\
                                        && NDCG& 0.126   & 0.1707& 0.1710  & 0.2103    &  0.2291  &\bf{0.2261} & 0.2245        & 6.75\%  \\ \hline

\hline
\hline
\end{tabular}
\caption{HR@10 and NDCG@10 comparisons of different methods w.r.t. the number of predictive factors}
\label{tab:wei2}
\end{table*}

\section{Experiments}
In this section, we conducted experiments to show the effectiveness of our proposed models. Moreover, extensive experiments were conducted to analyze the performance with different experimental settings, such as the number of hidden layers, negative sampling ratio, size of {\it predictive factors}, and so on.

\begin{table}[] %!hbp
\centering
\begin{tabular}{c c c c c}
\toprule
Dataset & \#users & \#items & \#interactions &density \\ [3pt]
\midrule 
% ML100k   &  973   &  1,684   &  100,000   &  6.29\%  \\ [3pt]

ML1m   &  6,040   &  3,260   &  998,539   &  5.07\%  \\ [3pt]

Amusic   &  5,729   &  9,267   &  65,344   &  0.12\%  \\ [3pt]

% Amusic-t  &  7947   &  25,975   &  134,860   &  0.07\%  \\ [3pt]
\bottomrule
\end{tabular}
\caption{Data statistic on two real-world datasets}
\label{tab:wei1}
\end{table}

\subsection{Experimental Settings}
\noindent\textbf{Datasets}\quad We evaluated our models on two real-world datasets form different domains, each of which has been widely used in many previous works for evaluation: MovieLens 1M\footnote{The dataset is available at https://grouplens.org/datasets/movielens/1m/} (ML1m) and Amazon Digital Music (Amusic)\footnote{The dataset is available at http://jmcauley.ucsd.edu/data/amazon/}. For both datasets, we discarded users and items associated with less than 10 interations. Table 2 shows the statistics of our two datasets.

\noindent\textbf{Evaluation Protocols}\quad We adopted the {\it leave-one-out} evaluation to evaluate the performance of item recommendation. For both datasets, we held out the latest interaction as a test item and the second latest interaction as a validation item for every user. The remaining data is used for training.  Since it is too time-consuming to rank all items for every user during testing, following \cite{koren2008factorization,He2017NCF}, we randomly sampled 100 items that are not interacted by the user, ranking the test item among the sampled items. The ranking performance is evaluated by {\it Hit Ratio} (HR) and {\it Normalized Discounted Cumulative Gain} (NDCG) \cite{he2015trirank}. For both metrics, the results are based on the truncated ranked list at 10.

\noindent\textbf{Baselines}\quad We compared our proposed methods with the following baselines. We leave out the comparison with item-item models, such as CDAE\cite{Wu2016CDA}, because they lack of user models for personalization, which may cause performance difference.
\begin{itemize}
\item \textbf{ItemPop}. Items are ranked by the number of interactions. It is a non-personalized method that is widely used as the baseline for personalized methods.
\item \textbf{BPR} \cite{Rendle2009} is a pairwise ranking method which optimizes the matrix factorization model with a pairwise ranking loss.
\item \textbf{eALS} \cite{he2016fast} is a state-of-the-art matrix factorization method with square loss for collaborative filtering with implicit feedback. 
\item \textbf{NeuMF} \cite{He2017NCF} is a state-of-the-art neural network based collaborative filtering method with binary cross-entropy loss. For fair comparison, we employ the same embedding size, number of hidden layers, and size of predictive factors for NueMF and our models.
% \item deepMF  is a state-of-the-art neural network based matrix factorization method, as we have mentioned before. We compare with deepMF-ce with 2 layers since it performs best in the implicit feedback scenario.
\end{itemize}
\begin{figure*}[!htbp]
\centering
% \subfigure[Amusic-HR@K]{\includegraphics[scale=0.55]{se1.pdf}}
\includegraphics[scale=0.4]{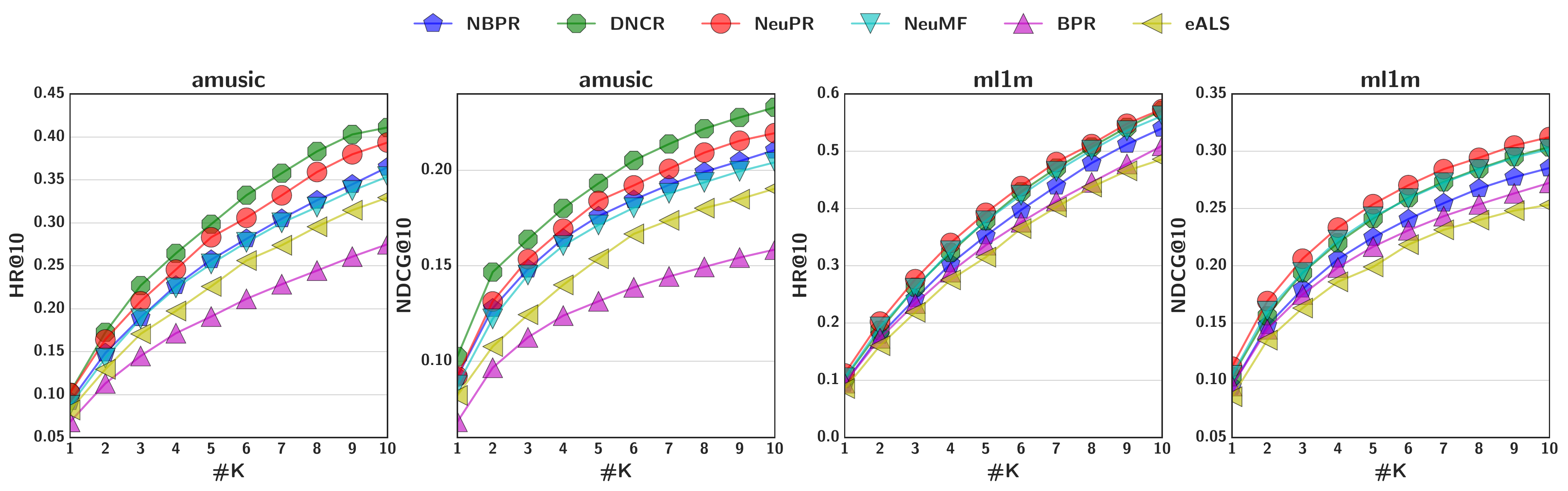}
% \subfigure[Amusic-NDCG@K]{\includegraphics[scale=0.55]{se1.pdf}}
% \subfigure[MovieLens-HR@K]{\includegraphics[scale=0.55]{hr2.pdf}}
% \subfigure[MovieLens-NDCG@K]{\includegraphics[scale=0.55]{ndcg2.pdf}}
\caption{Evaluation of Top-$K$ item recommendation where $K$ ranges from 1 to 10 on the two datasets}
\end{figure*}

\begin{figure*}[!htbp]
\centering
\includegraphics[scale=0.4]{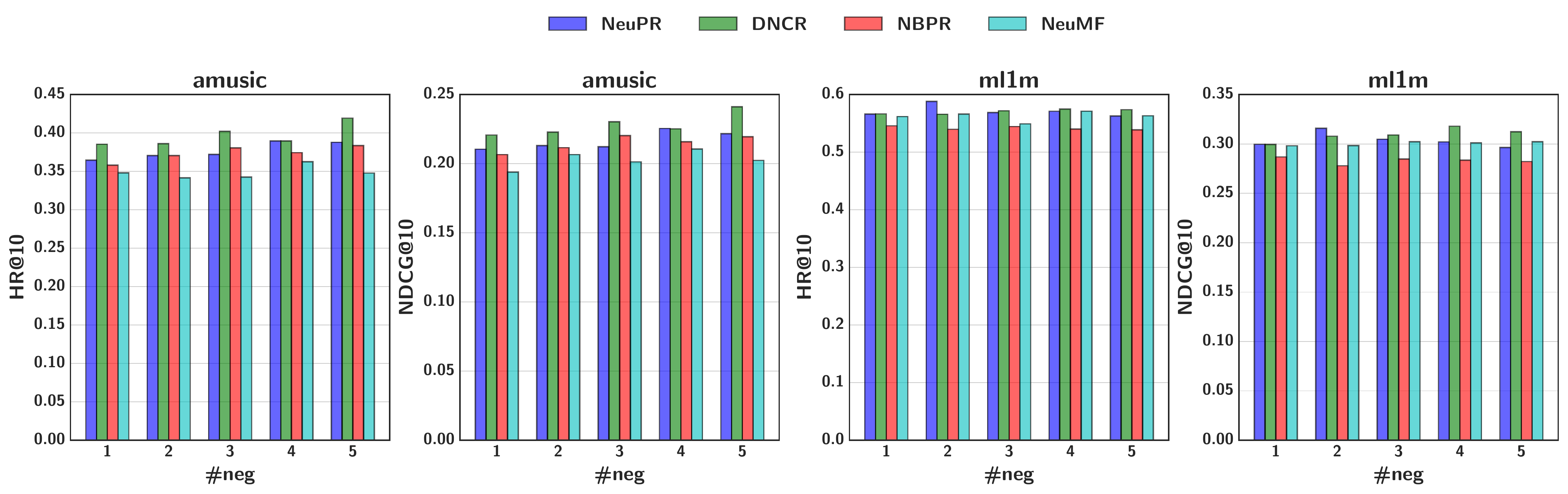}
% \subfigure[Amusic-HR@10]{\includegraphics[scale=0.55]{neg_hr1.pdf}}
% \subfigure[Amusic-NDCG@10]{\includegraphics[scale=0.55]{neg_ndcg1.pdf}}
% \subfigure[MovieLens-HR@10]{\includegraphics[scale=0.55]{neg_hr2.pdf}}
% \subfigure[MovieLens-NDCG@10]{\includegraphics[scale=0.55]{neg_ndcg2.pdf}}
\caption{Performance of NeuMF, NeuPR, DNCR and NBPR w.r.t. the number of negative samples per positive instance (factors=8).}
\end{figure*}

\noindent\textbf{Parameter Settings}\quad We implemented our proposed approaches based on keras\footnote{https://github.com/fchollet/keras}. For learning NCR, we randomly sampled one interaction for each user as the validation data and tuned hyperparameters on it. We varied the learning rate of [0.001, 0.0005, 0.0001], randomly initialized model weights with a Gaussian distribution (mean of 0 and standard deviation of 0.01), set the batch size to be 256, and chose Adam optimizer. For methods relying on negative samples, we sampled one negative instances per positive instances. And recall that in section 4.1 we term the last hidden layer of NCR as {\it predictive factors}. We conducted experiments to test the {\it predictive factors} of [8, 16, 24, 32, 64]. Without special mention, we employed four hidden layers for DNCR; for example, if the size of {\it predictive factors} is 8, then the neural network architecture of hidden layers is $96\rightarrow 32\rightarrow 16\rightarrow 8$, and the embedding size is 32. 
% For pre-training, we simply set $\alpha$ to 0.5.
\\\\\noindent\textbf{Performance Comparison}\quad Table 1 shows the results of comparison w.r.t. the number of {\it predictive factors}. For BPR and eALS, the size of {\it predictive factors} is equal to the dimensionality of latent factors. The table demonstrates the effectiveness of our proposed models, we can see that DNCR and NeuPR achieve the best performance in both metrics NDCG and HR on most cases. On both datasets, our models are able to outperform the state-of-the-art matrix factorization methods eALS and BPR by a considerable margin. Even in comparison with the strongest baseline, NeuMF, our NeuPR consistently outperforms it and can achieve relative improvements of 3.21\%$\sim$ 7.42\% by HR on MovieLens. On the same dataset, NeuPR achieves relative improvements of 2.60\%$\sim$ 5.72\% by NDCG in comparison with NeuMF. On Amusic, the corresponding improvements are 4.86\%$\sim$ 9.21\% and 3.06\%$\sim$ 8.92\%. If we take the best of all NCR models into consideration, NCR models can achieve relative improvements of 3.21\%$\sim$ 7.42\% by HR and 3.97\%$\sim$ 5.72\% by NDCG on the MovieLens dataset. While on the extremely sparse Amusic dataset, our models significantly outperform the strongest baseline NeuMF, the corresponding improvements are 5.12\%$\sim$ 14.53\% by HR and 7.1\%$\sim$ 12.86\% by NDCG (paired t-tests, p<0.01). 

%As can be seen, on MovieLens, NeuPR consistently outperforms other methods across $K$.

\indent Figure 3 shows the performance of Top-$K$ recommended lists where the number of recommended items ranges from 1 to 10. Here we employed {\it predictive factors} of 8 for all methods. And ItemPop is omitted due to its weak performance. DNCR generally achieves similar prediction accuracy in comparison with NeuMF, and their performance curves are so close that we can hardly make a distinction. On Amusic, NBPR achieves comparable prediction accuracy with NeuMF, DNCR and NeuPR better NeuMF by a considerable margin. On both dataset, neural-network-based methods outperform conventional matrix-factorization-based methods. 
The characteristic of datasets also have some influences on the results; on the extreme sparse Amusic dataset, the performance gaps between different methods are relatively large, while on the relatively dense MovieLens dataset, the performance gaps between different methods are relatively small.  From Table 1 and Figure 3, we can conclude that on dense dataset like MovieLens, NeuPR performs best, DNCR performs better than NBPR. While on the extreme sparse dataset like Amusic, DNCR performs best, NeuPR the second best and NBPR the worst. This indicates that on extreme sparse dataset, model's capacity is more important than it's ability to avoid overfitting. NeuPR's NBPR part may drag its performance on Amusic.
\\\\\noindent\textbf{Impact of Negative Sampling Ratio}\quad  We also conducted extensive experiments to compare the performance of NCR models with the strongest baseline NeuMF under different negative sampling ratio. Figure 4 shows the performance of NeuMF and NeuPR w.r.t. the number of negative samples per positive instance. As can be seen, on both datasets, NCR methods beat all other methods in terms of both metrics across different negative sampling ratio. Among three NCR methods, DNCR consistently outperforms the other two methods on Amusic; While on MovieLens, NeuPR performs the best. On Amusic, DNCR performs the best when the negative sampling ratio is 5 negative samples per positive sample. On MovieLens, NeuPR performs the best when the negative sampling ratio is 2 negative samples per positive samples.
% Another obvious result is that the performance of NeuPR does not change much when increasing negative sampling ratio, while the performance of NeuMF is relatively sensitive to negative sampling ratio. Moreover, the optimal negative sampling ratio of NeuMF is different on different dataset. Our empirical study shows that the optimal negative 
% sampling ratio of NeuMF is depended on the data. Intuitively, the reason why NeuMF is sensitive to negative sampling ratio is caused by the fact that NeuMF simply views interacted items as positive instances and the remaining as negative instances--- For each user, the number of interacted items is much less than that of non-interacted items, leading to the {\it Class Imbalance Problem} and demanding certain negative sampling ratio to alleviate this problem. While under our classification strategy the number of positive instances is equal to that of negative instances, which gives us hope to solve the {\it Class Imbalance Problem}. This has been confirmed by the fact that our model is not sensitive to negative sampling ratio. This means our models can get rid of the troublesome detail of tuning negative simpling ratio. In practice, sampling one negative instance per positive instance is sufficient to acquire good performance for NCR. 
\\\\\noindent\textbf{Training Loss}\quad To compare NeuPR, DNCR and NBPR more clearly, we further investigate the training loss (averaged over all instances) of NCR methods of each iteration on the two datasets. For fair comparison, we use learning rate of 0.0005, negative sampling ratio of 1 and report the training loss within 100 iterations. As can be seen in Figure 5, NeuPR achieves the lowest training loss on both datasets. However, lower training loss does not always means higher performance. For example, on Amusic, DNCR has the highest training loss while at the same time it achieves the best performance. 
\begin{figure}[!htbp]
\centering
\includegraphics[scale=0.33]{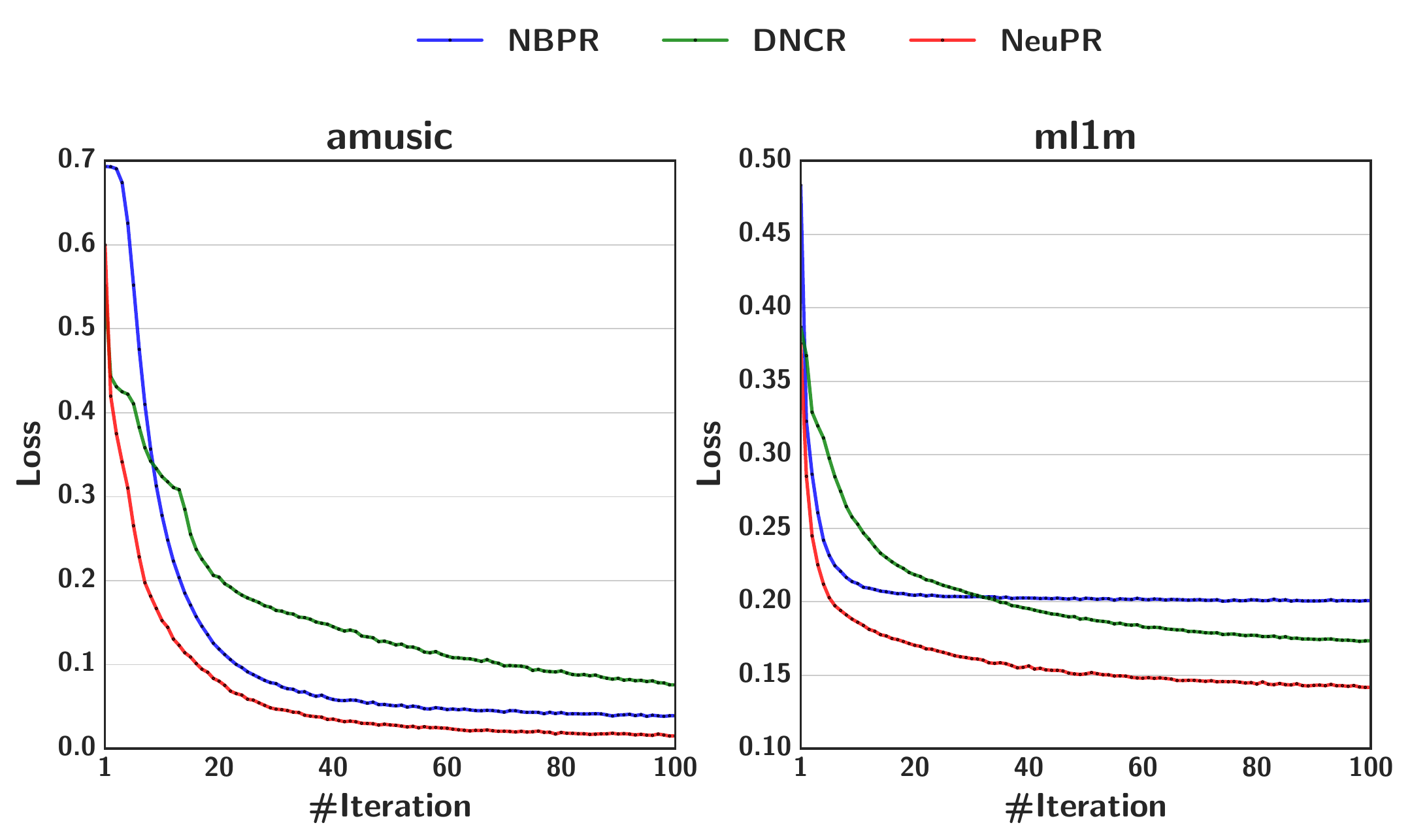}
\caption{Evaluation of pre-training w.r.t. $\alpha$ , where $\alpha$ ranges from 0.0 to 1.0.}
\end{figure}
\\\\
\noindent\textbf{Impact of Depth of Layers in DNCR} We conducted extensive experiments to investigate DNCR with different number of hidden layers. The results are shown in Figure 6. Here DNCR1 means DNCR with 1 hidden layers, and other DNCR notations have similar meaning.
\begin{figure}[!htbp]
\centering
\includegraphics[scale=0.33]{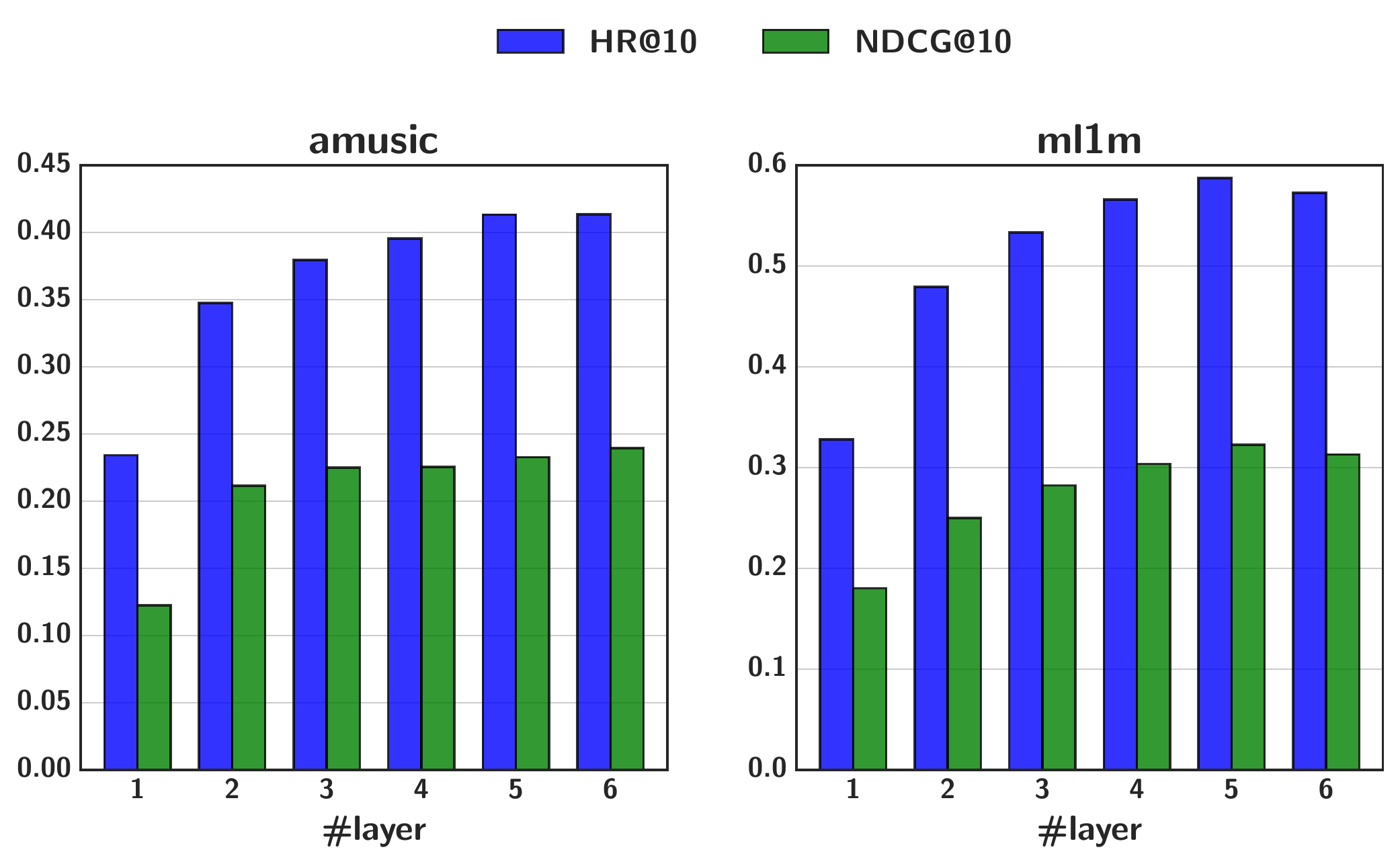}
\caption{Evaluation of DNCR w.r.t. the number of hidden layers, where the number of hidden layers ranges from 1 to 6.}
\end{figure}
We evaluated the performance using the same number of {\it predictive factors} (8) for DNCR with two or larger number of hidden layers. As can be seen, DNCR with only one hidden layer (In this case, the hidden layer is a concatenation of input features) performs worst, it only performs slightly better than Itempop, and underperforms eALS and BPR by a huge margin. Although DNCR2 only has one more hidden layer than DNCR1, it performs far better than DNCR1. This result shows that simply concatenating latent vectors is insufficient to capture the interactions between latent factors. On both datasets, when the number of layers are smaller than 5, increasing the number of hidden layers brings better performance;  When we use DNCR with more than 5 hidden layers, the performance does not improve. 
% In conclusion, DNCR performs the best when using 5 hidden layers.
\\\\\noindent\textbf{Utility of Pre-training}
We conducted an extensive experiment to investigate the utility of pre-training for NeuPR with $\alpha=0.5$. Table 3 shows the performance of NeuPR with and without pre-training. As can be seen, NeuPR with pre-training consistently outperforms NeuPR without pre-training on Amusic. On MovieLens, pre-training achieves better performance in most cases, but not significantly. On the Amusic dataset, pre-training is able to improve the recommendation quality by a large margin. In general, pre-training is beneficial to recommendation quality. 
\begin{table}[!htb]
\centering
\label{tab4}
\begin{tabular}{|c|c|c|c|c|}
\hline
 NeuPR  & \multicolumn{2}{c|}{Without Pre-training} & \multicolumn{2}{c|}{With Pre-training} \\\hline
factors & HR@10            & NDCG@10            & HR@10             & NDCG@10             \\\hline\hline
\multicolumn{5}{|c|}{Ml1m}                                                          \\\hline
8       &   0.5661         &    0.2997         &  \textbf{0.5702}    &    \textbf{0.3024}          \\\hline
16      &   0.5692         &    \bf{0.3124}    &  \textbf{0.5750}    &    0.3085          \\\hline
24      &   0.5724         &    0.3104         &  \textbf{0.5776}    &    \textbf{0.3127}           \\\hline                  
32      &   0.5740         &    0.3096         &  \textbf{0.5793}    &    \textbf{0.3122}      \\\hline
64      &   0.5801         &    \bf{0.3158}    &  \textbf{0.5840}    &    0.3125               \\\hline\hline
\multicolumn{5}{|c|}{Amusic}                                                             \\\hline
8       &  0.3645           &  0.2124            &  \textbf{0.4083}   &   \textbf{0.2342}                 \\\hline
16      &  0.3707           &  0.2089            &  \textbf{0.4036}   &  \textbf{0.2365}                  \\\hline
24      &  0.3736           &  0.2166            &  \textbf{0.3992}   &  \textbf{0.2327}                  \\\hline
32      &  0.3758           &  0.2185            &  \textbf{0.4025}   &   \textbf{0.2380}                 \\\hline
64      &  0.3826           &  0.2245            &  \textbf{0.4125}   &   \textbf{0.2429}                 \\\hline
\end{tabular}
\caption{Impact of Pre-training}
\end{table}

\noindent\textbf{Conclusion and Future Work}
\\In this work we propose a novel general neural network based collaborative ranking framework for personalized ranking. We experimentally demonstrate the effectiveness of our novel pairwise classification strategy for recommendation. The results on two real-world datasets illustrate the effectiveness of our proposed three NCR instantiations NBPR, DNCR and NeuPR. In future, we will study how to solve the problem of information loss caused by concatenating latent vectors, and how to extend our proposed framework to incorporate auxiliary information to enrich latent features.
 % We are also interested in exploring attention mechanism for personalized ranking.

\noindent \textbf{Acknowledgement}

\noindent This research is supported by the National Natural Science Foundation of China (NSFC) No.61672449. We thank Dr. Weike Pan in Shenzhen university for helpful discussions on pairwise ranking techniques.

{
\bibliographystyle{ACM-Reference-Format}
\bibliography{ncr}
}

\end{document}